\documentclass[twoside,slac_one]{revtex4}
\usepackage{graphicx}
\usepackage{fancyhdr}
\usepackage{amsmath} 
\usepackage{bm}
\usepackage{amsxtra}
\usepackage{amssymb}
\usepackage{amsthm}
\usepackage{latexsym}
\usepackage{lscape}

\begin{document}

\title{Studies with Onia at LHCb}

%

\author{L. Li Gioi - for the LHCb collaboration}
\affiliation{CNRS - Laboratoire de Physique Corpusculaire de Clermont-Ferrand, France}

\begin{abstract}
The production of $c\bar{c}$ and $b\bar{b}$ is studied  in $pp$ collisions at $\sqrt{s} = 7$~TeV with the LHCb detector. 
The results of these measurements are compared with different theoretical models. 
Results and prospects are also shown for exotics: the mass measurement of the $X(3872)$ and the search of the $X(4140)$.
Most of the presented results are based on the 2010 LHCb dataset (35 pb$^{-1}$).
\end{abstract}

\maketitle

\thispagestyle{fancy}


\section{Introduction}

Heavy quarkonium production remains a challenging problem for the understanding of Quantum Chromodynamics (QCD). At the centre-of mass energies for proton-proton collisions at the
Large Hadron Collider, $c\bar{c}$ pairs are expected to be produced predominantly via Leading Order gluon gluon interactions, which can be computed using perturbative QCD, followed by the formation of
the bound charmonium states described by non perturbative models. Recent approaches make use of non-relativistic QCD factorization (NRQCD) which assumes a combination of the colour-singlet 
and colour-octet $c\bar{c}$ as it evolves towards the final bound state via the exchange of soft gluons~\cite{ref:intr1}. Next-to-Leading Order (NLO) QCD corrections in charmonium and bottomonium
production are also essential for the description of the experimental data~\cite{ref:intr2,ref:intr3}. Studies of 
$\psi(2S)$, $\Upsilon(1S)$ and $\chi_c$ production cross sections are presented.

The $X(3872)$ meson is an exotic meson discovered in 2003 by the Belle collaboration in the $B^\pm \to X(3872) K^\pm$, $X(3872) \to J/\psi \pi^+\pi^-$ decay chain~\cite{ref:intr4}. 
Its existence was confirmed by the CDF~\cite{ref:intr5}, D0~\cite{ref:intr6} and BaBar~\cite{ref:intr7} collaborations. The $X(3872)$ particle, together with several other new states subsequently observed
in the mass range 3.9-4.7 GeV/c$^2$, has led to a resurgence of interest in exotic meson spectroscopy~\cite{ref:intr8}.
Several properties of the $X(3872)$ have been measured; however, its nature is still uncertain and several models have been proposed. First, it is not excluded that the $X(3872)$ is a
conventional charmonium state with one candidate being the $\eta_{c2}(1D)$ meson~\cite{ref:intr8}. However, the mass of this state is predicted to be far below the observed $X(3872)$ mass. 
Given the proximity of the $X(3872)$ mass to the $D^{*0} \bar{D^0}$ threshold, one possibility is that the $X(3872)$ is a loosely bound deuteron-like $D^{*0} \bar{D^0}$ molecule, 
i.e. a ((uc) - (cu)) system~\cite{ref:intr8}. Another more exotic possibility is that the $X(3872)$ is a tetraquark state~\cite{ref:intr9}.

The CDF experiment has reported $>5 \ \sigma$ evidence for $X(4140)$ state (also referred to as
$Y(4140)$ in the literature) in 6.0 fb$^{-1}$ of $p\bar{p}$ data collected at Tevatron~\cite{ref:intr10}.
The relative rate was measured to be ${B(B^+ \to X(4140)K^+) \times B(X(4140) \to J/\psi \phi)}/{B(B^+ \to J/\psi \phi K^+)} = 0.149 \pm 0.039 \pm 0.024$. 
This observation has triggered wide interest among model builders of exotic hadronic states. It has been suggested that the $X(4140)$ could be a molecular state~\cite{ref:intr11,ref:intr12}, 
a tetraquark state~\cite{ref:intr13,ref:intr14}, a hybrid state~\cite{ref:intr15,ref:intr16} or a re-scattering effect~\cite{ref:intr17,ref:intr18}.

\section{\boldmath{$\psi(2S)$} Production Cross Section}

Two decay modes of the $\psi(2S )$ meson have been studied~\cite{ref:confPsi2S}: $\psi(2S ) \to \mu^+\mu^-$ and $\psi(2S ) \to J/\psi(\mu\mu)\pi^+\pi^-$. 
The differential cross-section for the inclusive $\psi(2S)$ production is computed as
\begin{equation}
 \frac{d\sigma}{dp_T}(p_T) = \frac{N_{\psi(2S)}(p_T)}{L_{int} \epsilon(p_T) B(\psi(2S) \to f) \Delta p_T}
\end{equation}
where $N_{\psi(2S)}(p_T)$ is the number of observed $\psi(2S )$ decays, $L_{int}$ is the integrated luminosity, $\epsilon(p_T)$ is the total detection efficiency including acceptance 
effects, $B(\psi(2S) \to f)$ is $B(\psi(2S) \to J/\psi\pi^+\pi^-) B(J/\psi \to \mu^+\mu^-)$ for $\psi(2S ) \to J/\psi(\mu\mu)\pi^+\pi^-$ and the dielectron branching ratio 
($B(\psi(2S) \to e^+e^-)$) for $\psi(2S ) \to \mu^+\mu^-$ and $\Delta p_T =$ 1 GeV/c is the bin size. The dielectron branching ratio is used, assuming lepton universality, since it has a much 
smaller error than the dimuon one (2.2\% and 10\% respectively). In order to estimate the number of $\psi(2S)$ signal events, a fit is performed independently in each $(p_T)$ bin. 
\begin{figure}[t]
\centering
\includegraphics[width=80mm]{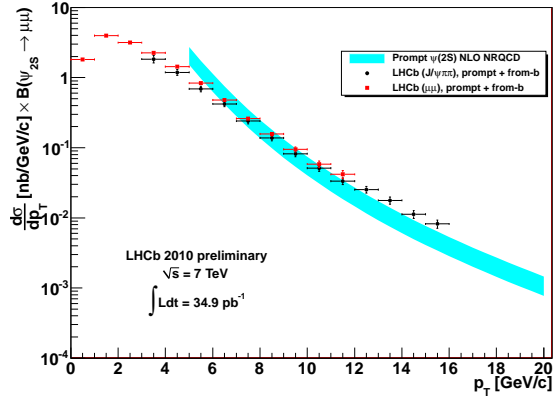}
\caption{Comparison of the LHCb results for the differential production cross-section of $\psi(2S)$ with the predictions for prompt production by a NLO NRQCD model~\cite{ref:Psi2S_1}. LHCb data include
also $\psi(2S)$ from b.} 
\label{Fig:Psi2S_1}
\end{figure}
In Fig.~\ref{Fig:Psi2S_1} a comparison between the measurements presented of the differential $\psi(2S)$ production cross-section multiplied by the $\psi(2S ) \to \mu^+\mu^-$ branching ratio and a recent 
theory prediction~\cite{ref:Psi2S_1} for prompt $\psi(2S)$ in the rapidity range $2 \leq y= \frac{1}{2}ln\frac{E+p_z}{E-p_z} \leq 4.5$ is shown. Here $E$ and $p_z$ are the $\psi(2S )$ energy and 
momentum in the $z$ direction measured in the pp centre-of-mass frame; the z-axis is defined along the beam axis in the LHCb frame, oriented from the VELO to the Muon Detector.
The differential cross-section in~\cite{ref:Psi2S_1} has been evaluated for the $\psi(2S)$ 
prompt production at the LHC at next-to-leading order in non-relativistic QCD, including color-singlet and color-octet contributions.
The integrated cross-section in the full range of $p_T$ and $y$ respectively to $\psi(2S) \to \mu^+\mu^-$ and $\psi(2S) \to J/\psi\pi^+\pi^-$ is found to be:
\begin{eqnarray}
\sigma(0 < p_T \leq 12 \ \mbox{GeV/c}, 2 < y \leq 4.5) & = & 1.88 \pm 0.02 \pm 0.31^{+0.25}_{-0.48} \ \mbox{$\mu$b}  \nonumber \\
\sigma(3 < p_T \leq 16 \ \mbox{GeV/c}, 2 < y \leq 4.5) & = & 0.62 \pm 0.04 \pm 0.12^{+0.07}_{-0.14} \ \mbox{$\mu$b}
\end{eqnarray}
where the first uncertainty is statistical, the second is systematic and the third is the uncertainty due to the unknown polarization.

\section{\boldmath{$\Upsilon(1S)$} Production Cross Section}

\begin{figure}[t]
\centering
\includegraphics[width=80mm]{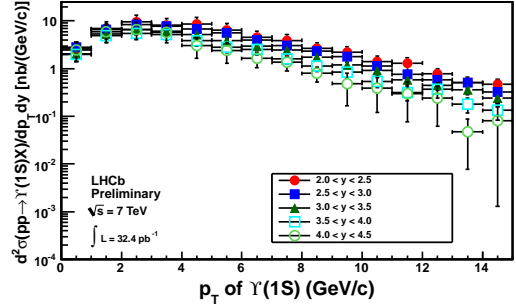}
\caption{Differential $\Upsilon(1S)$ production cross-section as a function of $p_T$ in bins of rapidity.} 
\label{Fig:Y1S_1}
\end{figure}

\begin{figure}[t]
\centering
\includegraphics[width=80mm]{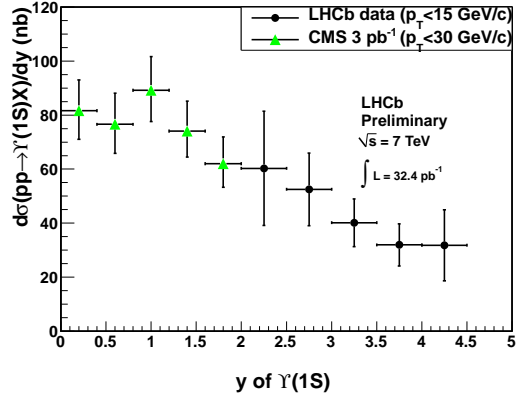}
\caption{Differential $\Upsilon(1S)$ production cross-section as a function of $y$ integrated over $p_T$ as measured by the CMS~\cite{ref:Y1S_2} and LHCb experiments.} 
\label{Fig:Y1S_2}
\end{figure}

$\Upsilon(1S)$ meson has been studied in the decay mode $\Upsilon(1S) \to \mu^+\mu^-$~\cite{ref:confY1S}. The double differential cross-section for the inclusive $\Upsilon(1S)$ production is computed as
\begin{equation}
 \frac{d^2\sigma}{dp_T dy}(p_T,y) = \frac{N_{\Upsilon(1S)}(p_T,y)}{L_{int} \epsilon(p_T,y) B(\Upsilon(1S) \to \mu^+\mu^-) \Delta p_T \Delta y}
\end{equation}
where $N_{\Upsilon(1S)}(p_T,y)$ is the number of observed $\Upsilon(1S) \to \mu^+\mu^-$ decays, $\epsilon(p_T,y)$ is the total detection efficiency including acceptance effects, $L_{int}$ is the integrated 
luminosity, $B(\Upsilon(1S) \to \mu^+\mu^-)$ is the branching fraction, and $\Delta p_T \Delta y = 1 \times 0.5$ GeV/c is the rapidity and $p_T$ bin sizes. 
In order to estimate the number of $\Upsilon(1S)$ signal events, a fit is performed independently in each of the 15 $p_T$ times 5 $y$ bins.
The double differential cross-section as a function of $p_T$ and $y$ is shown in Fig.~\ref{Fig:Y1S_1}. The integrated cross-section in the full range of $y$ and $p_T$ is found to be
\begin{equation}
 \sigma(pp \to \Upsilon(1S)X; p_T(\Upsilon(1S)) < 15 \ \mbox{GeV/c}; 2 < y(\Upsilon(1S)) < 4.5) = 108.3 \pm 0.7 ^{+30.9}_{-25.8} \ \mbox{nb}
\end{equation}
where the first uncertainty is statistical, and the second systematic. The latter includes ${}^{+18.8}_{-7.9}$ nb from the unknown polarization, $\pm 10.8$ nb from the luminosity determination and 
$\pm 22.0$ nb from other sources. The integrated cross-section is about a factor 100 smaller than the integrated $J/\psi$  cross-section in the identical $y$ and $p_T$ region~\cite{ref:Y1S_1},
and a factor three smaller than the integrated $\Upsilon(1S)$ cross-section in the central region $|y| < 2$ as measured by CMS~\cite{ref:Y1S_2}.
The CMS and the LHCb measurements agree; the difference in the integrated cross section results is due to the different rapidity ranges.
Fig.~\ref{Fig:Y1S_2} shows the comparison of the LHCb cross-section measurement in bins of $y$ integrated over $p_T$ with the same measurement of CMS, in the $y$ ranges covered 
by the two experiments.

\section{\boldmath{$\sigma(\chi_{c2})/\sigma(\chi_{c1})$} Production Cross Sections Ratio}

\begin{figure}[t]
\centering
\includegraphics[width=80mm]{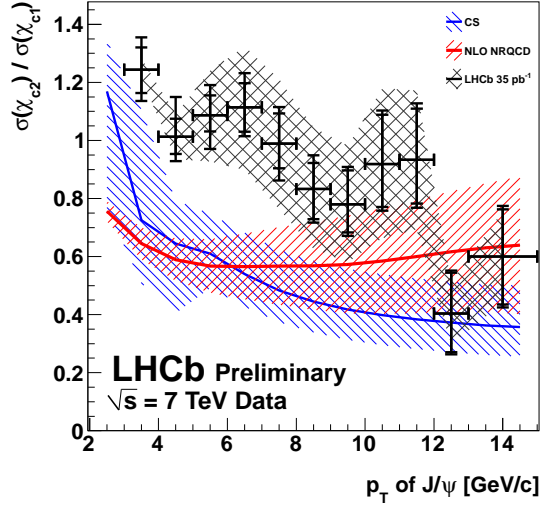}
\caption{The ratio $\sigma(\chi_{c2})/\sigma(\chi_{c1})$ in bins of  in the range of $J/\psi \ p_T$ : [3; 15] GeV/c. The internal error bars correspond to the statistical error on the $\chi_{c1}$ and $\chi_{c2}$
yields; the external error bars include the contribution from the systematic uncertainties (apart from the polarization). The shaded area around the data points (black) shows the maximum effect of the unknown 
$\chi_c$ polarization on the result. The two other bands correspond to the ChiGen MC generator theoretical prediction~\cite{ref:Chic_1} (in blue) and
NLO NRQCD~\cite{ref:Chic_2} (in red).} 
\label{Fig:chic_1}
\end{figure}

$\chi_{c1}$ and $\chi_{c2}$ have been reconstructed in the final state $J/\psi \gamma$. The production cross-section ratio of the $\chi_{c2}$ and $\chi_{c1}$ states is measured~\cite{ref:confChic} using
\begin{equation}
\frac{\sigma (\chi_{c2})}{\sigma (\chi_{c1})} = \frac{N_{\chi_{c2}}}{N_{\chi_{c1}}}\cdot
\frac{\epsilon^{\chi_{c1}}_{J/\psi} \epsilon^{\chi_{c1}}_{\gamma} \epsilon^{\chi_{c1}}_{sel}}{\epsilon^{\chi_{c2}}_{J/\psi} \epsilon^{\chi_{c2}}_{\gamma} \epsilon^{\chi_{c2}}_{sel}} 
\cdot \frac{B(\chi_{c1}\rightarrow J/\psi \gamma)}{B(\chi_{c2}\rightarrow J/\psi \gamma)}
\end{equation}
where $B(\chi_{c1}\rightarrow J/\psi \gamma)$ and $B(\chi_{c2}\rightarrow J/\psi \gamma)$ are the $\chi_{c1}$ and $\chi_{c2}$ branching ratios to the final state $J/\psi \gamma$,
$\epsilon^{\chi_{c2}}_{J/\psi}$ ($\epsilon^{\chi_{c2}}_{\gamma}$) is the efficiency to reconstruct and select a $J/\psi$ ($\gamma$) from $\chi_{c}$ decay and $\epsilon^{\chi_{c}}_{sel}$ is the efficiency 
to select the $\chi_{c}$ candidate. 
The measurement method consists of extracting the two $N_{\chi_{ci}}$ yields (for $\chi_{c1}$ and $\chi_{c2}$) from an unbinned maximum likelihood fit to $\Delta M = M(\chi_{c})-M(J/\psi)$ mass difference 
distribution.
With the ratio of the resolution parameters and the mass differences fixed, a fit is then performed to the data in the full $J/\psi \ p_T$ range $J/\psi \ p_T \in [3,15]$ GeV/c in order to extract the 
resolution scale $\sigma_{res}(\chi_{c1})$. Here, the sample is subdivided into candidates with converted (after the magnet) and non-converted photons, in order to account for the different calorimeter 
resolution in the two cases. Converted photons are identified by the presence of activity in the scintillator pad detector at the entrance to the calorimeter system. Photons that convert before the magnet 
have a low probability to be reconstructed, because either one or both electrons are swept out of the detector acceptance by the magnetic field, and are not considered here.
The resolution scales, $\sigma_{res}(\chi_{c1})$, are measured to be $22.8 \pm 1.1$ MeV/c$^2$ and $18.4 \pm 0.4$ MeV/c$^2$ for converted and non converted candidates, respectively.
The fit is then performed in bins of $J/\psi \ p_T$.  
For each bin the value of $\sigma_{res}(\chi_{c1})$ is fixed to the value extracted from the fit to the full range $J/\psi \ p_T$.
The presence of $\chi_{c}$ polarized states would modify the efficiencies calculated from the Monte Carlo. In order to take into account possible polarization scenarios, the relevant combination of
weights has to be taken into account.
The results from the not-converted and converted samples are combined by sampling the combination of the statistical and uncorrelated systematic uncertainties using a toy Monte Carlo. 
Finally, the correlated systematic uncertainty from the branching ratio is calculated using the combined central values for $\sigma(\chi_{c2})/\sigma(\chi_{c1})$. The preliminary
result for the ratio of the prompt $\chi_{c2}$ to $\chi_{c1}$ production cross-sections as a function of $J/\psi \ p_T$ is given in Fig.~\ref{Fig:chic_1}.
Comparisons to the theory predictions from the ChiGen MC generator~\cite{ref:Chic_1} and from the NLO NRQCD calculations~\cite{ref:Chic_2}, in the rapidity range [2; 4.5], are also shown in
the figure. 
Fig.~\ref{Fig:chic_1} also shows the maximum effect of the unknown $\chi_{c}$ polarization on the result, shown as a black shaded area around the data points. The upper limit of the shaded area
corresponds to the spin state ( $\chi_{c1}$ : $m_J$ = 1;  $\chi_{c2}$ : $m_J$ = 2) and the lower limit corresponds to the spin state ( $\chi_{c1}$ : $m_J$ = 0; $\chi_{c2}$ : $m_J$ = 0). 
The results are broadly in agreement at high $J/\psi \ p_T$ with the colour singlet model, however, they are not yet precise enough to rule out sizeable colour octet terms. There
are indications of a discrepancy in the mid to low $J/\psi \ p_T$ region. This may be explained by a more complete modeling of the transition from the high to low $J/\psi \ p_T$ regions, which is
sensitive to non-perturbative effects and/or sizeable higher-order perturbative corrections.

\section{\boldmath{$X(3872)$} Mass Measurement}

Inclusively produced $X(3872)$ mesons are reconstructed and selected in the $X(3872) \to J/\psi \pi^+\pi^-$, $J/\psi \to \mu^+\mu^-$ decay mode~\cite{ref:confXmass}. A momentum scale calibration 
is performed to account for a mixture of effects related to imperfections in the knowledge of the magnetic field map and of the alignment of the tracking system.

\subsection{Momentum Scale Calibration}

The momentum scale is calibrated using a large sample of $J/\psi \to \mu^+\mu^-$ decays. This calibration, gives an overall scale factor that is applied to all measurements of particle momenta.
The calibration is checked to be also valid for two-body decays of the $\Upsilon(1S)$, $D^0$
and $K^0$. In addition, its effect on $\psi(2S) \to J/\psi \pi^+\pi^-$, which have kinematics similar to the $X(3872) \to J/\psi \pi^+\pi^-$, is studied. The uncalibrated
value of the $\psi(2S)$ mass in data is $3685.94 \pm 0.06(stat)$ MeV/c$^2$. After the momentum scale calibration it becomes $3686.12 \pm 0.06(stat)$ MeV/c$^2$, in good agreement with the PDG
value of $3686.09 \pm 0.04$ MeV/c$^2$. The measured  $J/\psi$ mass after alignment and calibration is checked to be stable over
the whole 2010 data-taking period.

\subsection{Results}
\label{sec_Xmass_res}

\begin{figure}[t]
\centering
\includegraphics[width=80mm]{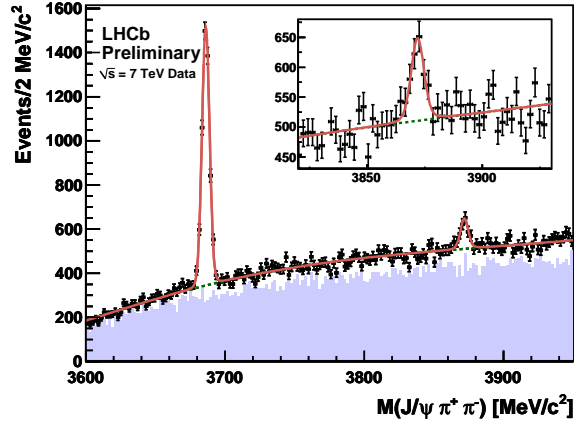}
\caption{Invariant mass distribution of $J/\psi \pi^+\pi^-$ (black points with statistical error bars) and same-sign $J/\psi \pi^\pm\pi^\pm$ (blue filled histogram) candidates. 
The red curve is the result of the fit described in the text. The insert shows a zoom of the region around the $X(3872)$ mass.} 
\label{Fig:Xmass_1}
\end{figure}

The masses of the $\psi(2S)$ and $X(3872)$ are determined from an extended unbinned maximum likelihood fit of the reconstructed $J/\psi \pi^+\pi^-$ mass in the interval 
$3.6 < M_{J/\psi \pi^+\pi^-} < 3.95$ GeV/c$^2$. The functional form of the background is studied using the same-sign pion events.
The $\psi(2S)$ and $X(3872)$ signals are each described with a Voigt function defined as the convolution of a non-relativistic Breit-Wigner with a Gaussian function.
The intrinsic width of the  $\psi(2S)$ is fixed to the PDG value, $\Gamma_{\psi(2S)} = 0.317$ MeV/c$^2$. The $X(3872)$ intrinsic width is poorly known. The BaBar~\cite{ref:Xmass_1} and Belle~\cite{ref:intr4} collaborations
have published $90\%$ confidence level limits of $\Gamma_{X(3872)} < 3.3$ MeV/c$^2$ and $\Gamma_{X(3872)} < 2.3$ MeV/c$^2$, respectively. The results of the Belle and BaBar analyses have
been combined by CDF to yield $\Gamma_{X(3872)} = 1.3 \pm 0.6$ MeV/c$^2$~\cite{ref:Xmass_2}. For these studies, the procedure adopted in~\cite{ref:Xmass_2} has been adopted and the natural width has been fixed to this value in the default fit.
Fig.~\ref{Fig:Xmass_1} shows the invariant $J/\psi \pi^+\pi^-$ mass distributions for opposite-sign (black points) and same-sign (blue filled histogram) candidates. Clear signals for both the $\psi(2S)$ and the
$X(3872)$ can be seen. From a comparison of fits with and without the $X(3872)$ component the statistical significance of the $X(3872)$ signal is estimated to be $9 \ \sigma$.  
The uncertainties on the parameters reported by the fit are in good agreement with expectations based on toy Monte Carlo studies. The preliminary LHCb result 
\begin{equation}
 M_{X(3872)} = 3871.96 \pm 0.46 \pm 0.10 \ \mbox{MeV/c$^2$}
\end{equation}
is in good agreement with the published measurements~\cite{ref:intr4, ref:Xmass_1, ref:Xmass_2, ref:intr5} and with their average, $3871.56 \pm 0.22$ MeV/c$^2$.
The new world average value including LHCb measurement, $3871.63 \pm 0.20$ MeV/c$^2$, is consistent within uncertainties with the sum of the $D^0$ and $D^{*0}$ masses, 
$3871.79 \pm 0.29$ MeV/c$^2$, computed from the results of the global PDG fit of the charm meson masses.

\section{\boldmath{$X(3872)$} Production Cross Section}

The product of the inclusive production cross-section $\sigma(pp \to X(3872) + ...)$ by the $X(3872) \to J/\psi \pi^+\pi^-$ branching fraction is computed as~\cite{ref:confXCrSec}
\begin{equation}
 \sigma(pp \to X(3872) + ...) \times B(X(3872) \to J/\psi \pi^+\pi^-) = \frac{N^{corr}_{X(3872)}}{\eta_{tot} \times L_{int} \times B(J/\psi \to \mu^+\mu^-)}
\end{equation}
where $N^{corr}_{X(3872)}$ is the efficiency-corrected yield of $X(3872) \to J/\psi(\mu^+\mu^-) \pi^+\pi^-$ signal decays, $\eta_{tot}$ is a multiplicative factor to the efficiency that accounts 
for known differences between the data and simulation, $L_{int}$ is the integrated luminosity, and the notation $B(...)$ is used for branching fractions.
The $X(3872)$ signal yield is determined from an extended, unbinned maximum likelihood fit of the reconstructed $J/\psi \pi^+\pi^-$ mass in the interval $3.82 < M_{J/\psi \pi^+\pi^-} < 3.95$ GeV/c$^2$. 
The fitting function is the same used for the $X(3872)$ mass measurement. 
The mass resolution is fixed in this analysis to
\begin{equation}
 \sigma(X(3872)) = \sigma(\psi(2S)) \frac{\sigma^{MC}_{X(3872)}}{\sigma^{MC}_{\psi(2S)}} = 3.26 \pm 0.10 \ \mbox{MeV/c$^2$}
\end{equation}
where $\sigma(\psi(2S)) = 2.48 \pm 0.08$ MeV/c$^2$ is the mass resolution fitted in data for the kinematically similar decay $\psi(2S) \to J/\psi \pi^+\pi^-$, and where $\sigma^{MC}_{X(3872)}$ and
$\sigma^{MC}_{\psi(2S)}$ are the mass resolutions determined from fully simulated Monte Carlo events. 
As in section~\ref{sec_Xmass_res}, the natural width is fixed to 1.3 MeV/c$^2$. The effect of fixing parameters that are not perfectly known is investigated as part 
of the systematic studies. The $X(3872)$ signal yield is approximately $68\%$ of that reported in~\cite{ref:confXmass} due to the additional trigger requirements and fiducial cuts on $y$ and $p_T$.
Applying the procedure discussed above, and using the $p_T-$ and $y-$dependent efficiency from simulation, the efficiency-corrected signal yield is $N^{corr}_{X(3872)} = 9597 \pm 2217$, where the 
quoted uncertainty is statistical. Two factors enter into the determination of the correction factor $\eta_{tot}$. The first is a factor $1.024 \pm 0.011$ which accounts for differences in the efficiency 
of the muon identification in the data and simulation. The second is a factor of $0.96 \pm 0.02$ that accounts for observed differences in the efficiency of global event cuts applied in the trigger.
Multiplying these values together gives $\eta_{tot} = 0.983 \pm 0.023$. The uncertainty on this number is taken into account in the estimation of the systematic uncertainty. The preliminary measurement is
\begin{equation}
 \sigma(pp \to X(3872) + ...) \times B(X(3872) \to J/\psi \pi^+\pi^-) = 4.74 \pm 1.10 \pm 1.01 \ \mbox{nb}
\end{equation}
where $\sigma(pp \to X(3872) + ...)$ is the cross-section for producing an $X(3872)$ particle in $pp$ collisions at $\sqrt{s} = 7$~TeV (either promptly or from the decay of other particles) with a
transverse momentum between 5 and 20 GeV/c and a rapidity between 2.5 and 4.5.

\section{Search of the \boldmath{$X(4140)$}}

\begin{figure}[t]
\centering
\includegraphics[width=80mm]{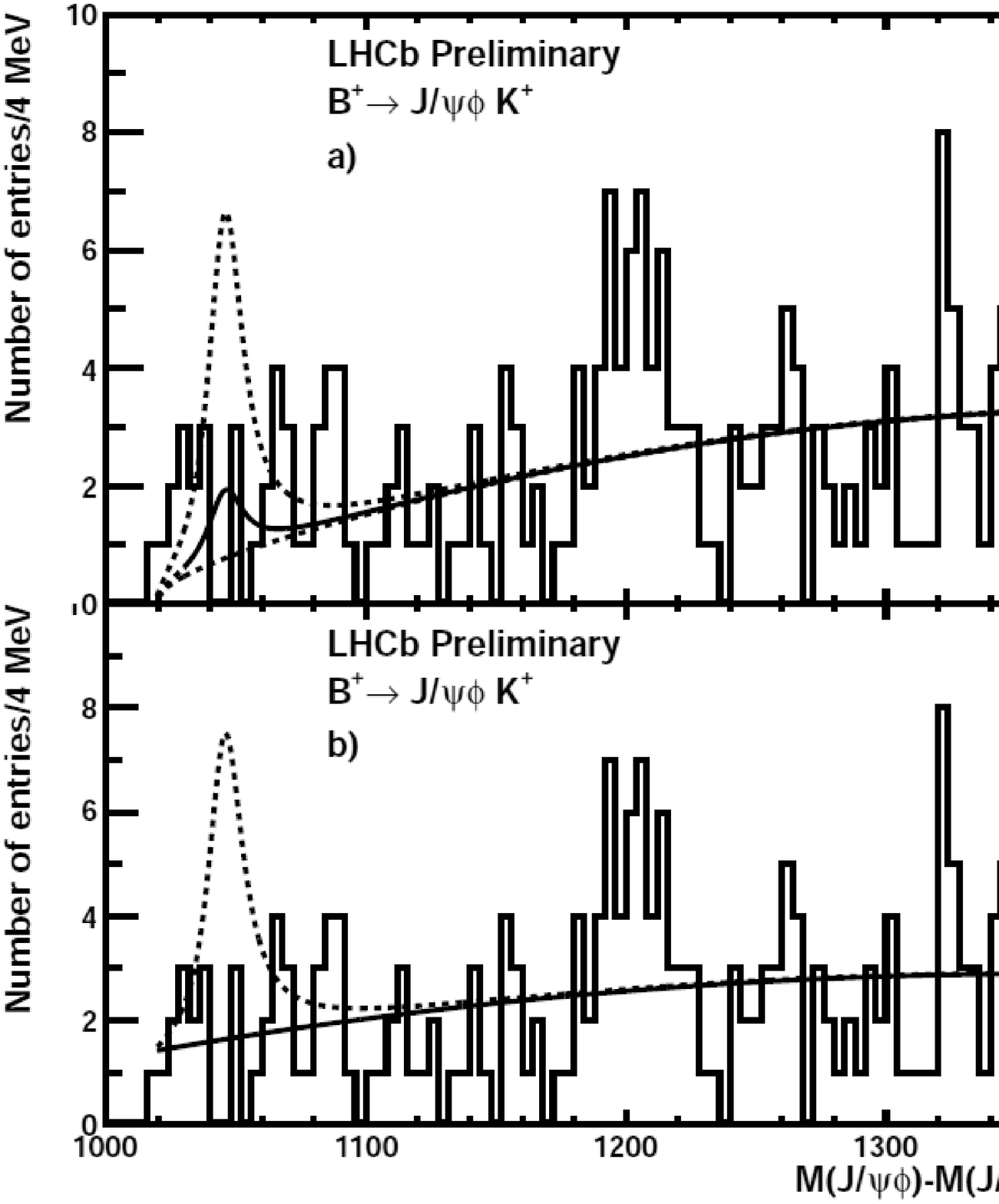}
\caption{Fit of $X(4140)$ signal on top of smooth background to $M(J/\psi \phi) - M(J/\psi)$ for the $B^+ \to J/\psi \phi K^+$ data. The solid line represents the result of the fit to LHCb data. The dashed
line on top illustrates the expected signal amplitude from the CDF results as explained in the text. The top and bottom plots differ by the type of the background function used in the fit: a)
efficiency-shaped three-body phase-spaced; b) quadratic polynomial.} 
\label{Fig:X4140_1}
\end{figure}

The $X(4140)$ has been studied using a data sample of approximately 0.376 fb$^{-1}$~\cite{ref:confX4140} selecting $B^+ \to J/\psi \phi K^+$, $\phi \to K^+K^-$.
No narrow structure has been seen near the threshold in $M(J/\psi \phi) - M(J/\psi)$ distribution as shown in Fig.~\ref{Fig:X4140_1}. In the CDF analysis, they fit their data with spin 0 
relativistic Breit-Wigner function on top of three-body phase-space, all smeared with the detector resolution~\cite{ref:intr10}. To quantify the disagreement with CDF the same function has been used. The efficiency 
dependence is extracted from the MC simulations and applied as a correction to the three-body phase-space function. Mass and width of $X(4140)$ peak are fixed to the values obtained by the CDF collaboration. 
The mass difference resolution is determined from the $B^+ \to X(4140)K^+$ MC simulation. 
The fit to $M(J/\psi \phi) - M(J/\psi)$ distribution gives a $X(4140)$ amplitude of $6.9 \pm 4.7$ events (Fig.~\ref{Fig:X4140_1}a) and has a confidence level (CL) of $3\%$. 
When using a quadratic polynomial instead of three-body phase-space function for the background, the preferred value of the $X(4140)$ amplitude is zero (it is restricted not to go below zero), with a
positive error of 3 events. This fit is shown in Fig.~\ref{Fig:X4140_1}b and has a confidence level of $11\%$.
Using the $B^+ \to J/\psi \phi K^+$ yield multiplied by this efficiency ratio ($B^+ \to X(4140)K^+$, $X(4140) \to J/\psi \phi$)/($B^+ \to J/\psi \phi K^+$) and multiplied by the CDF value for
$B(B^+ \to X(4140)K^+)/B(B^+ \to J/\psi \phi K^+)$~\cite{ref:intr10}, leads to a prediction of observed $35 \pm 9 \pm 6$ events, where the first uncertainty is statistical from the CDF data and the second is systematic 
including both CDF and LHCb contributions. The central value of this estimate is illustrated in Fig.~\ref{Fig:X4140_1}. The CDF result disagrees by three standard deviations with the fit to LHCb data using 
the polynomial background. The disagreement is 2.4 $\sigma$ when the efficiency-corrected three-body phase-space background shape is used.

\section{Exclusive Dimuon Production}

Exclusive particle production in proton-proton collisions are elastic processes in which the protons remain intact, $p + p \to p + X + p$, and the additional particles are created through photon and/or gluon 
propagators. In the case of purely photon propagators, this is an electromagnetic process which can be theoretically calculated with high accuracy. When gluons are involved, these processes constitute an 
important testing ground for QCD, since the object that couples to the proton must be colourless. Thus the pomeron (two gluon states) or possibly an odderon (three gluons), predicted in QCD but never 
unambiguously observed, can be studied in a clean experimental environment. The cleanest experimental manifestation of these occur in final states containing two protons and two muons or in events containing 
two protons, two muons and a photon. The former can be produced in the diphoton process giving a continuous dimuon invariant mass spectrum, or in the photon-pomeron process which can produce $\phi$, $J/\psi$,
$\psi(2S)$, $\Upsilon(1S)$, $\Upsilon(2S)$ or $\Upsilon(3S)$ which decay to two muons. The latter is a signal for double pomeron exchange which produces $\chi_c$($\chi_b$) that decay to $J/\psi$ ($\Upsilon$) 
plus a photon. The final state protons are only marginally deflected, go down the beam-pipe, and remain undetected. The experimental signal therefore in LHCb is a completely empty event except for two muons 
and possibly a photon. However, because LHCb is not hermetic, there will be sizeable backgrounds from non-elastic processes where the other particles travel outside the detector acceptance.

The cross-section $\sigma$, is calculated~\cite{ref:confExcl} from the number, $N$, of selected events having corrected for efficiency, $\epsilon$, and purity, $p$, and dividing by the luminosity, $L$, via 
$\sigma = (p N)/(\epsilon L)$. The efficiency for selecting the events has been determined from simulation. 
The measured cross-sections are:
\begin{eqnarray}
\sigma_{J/\psi \to \mu^+\mu^−}(2 < \eta_{\mu^+}, \eta{\mu^−} < 4.5) & = & 474 \pm 12 \pm 51 \pm 92 \ \mbox{pb} \nonumber \\
\sigma_{\psi(2S) \to \mu^+\mu^−}(2 < \eta_{\mu^+}, \eta{\mu^−} < 4.5) & = & 12.2 \pm 1.8 \pm 1.3 \pm 2.4 \ \mbox{pb} \nonumber \\
\sigma_{\chi_{c0} \to J/\psi\gamma \to \mu^+\mu^−\gamma}(2 < \eta_{\mu^+}, \eta{\mu^−}, \eta_{\gamma} < 4.5) & = & 9.3 \pm 2.2 \pm 3.5 \pm 1.8 \ \mbox{pb} \\
\sigma_{\chi_{c1} \to J/\psi\gamma \to \mu^+\mu^−\gamma}(2 < \eta_{\mu^+}, \eta{\mu^−}, \eta_{\gamma} < 4.5) & = & 16.4 \pm 5.3 \pm 5.8 \pm 3.2 \ \mbox{pb} \nonumber \\
\sigma_{\chi_{c2} \to J/\psi\gamma \to \mu^+\mu^−\gamma}(2 < \eta_{\mu^+}, \eta{\mu^−}, \eta_{\gamma} < 4.5) & = & 28.0 \pm 5.4 \pm 9.7 \pm 5.4 \ \mbox{pb} \nonumber \\
\sigma_{pp \to p\mu^+\mu^-p}(2 < \eta_{\mu^+}, \eta{\mu^−} < 4.5,M_{\mu^+\mu^-}> 2.5 \ \mbox{GeV/c$^2$}) & = & 67 \pm 10 \pm 7 \pm 15 \ \mbox{pb} \nonumber 
\end{eqnarray}

where the first uncertainty is statistical, the second is systematic, and the third comes from the estimate of the luminosity. Note that these numbers are cross-section times the branching ratio into 
the final state of interest, and all final state particles are required to be between pseudorapidities of 2 and 4.5. The cross sections are quoted in a limited pseudorapidity range as the models
under consideration have pseudorapidity dependence.

\section{Summary}

LHCb performed many analysis of the quarkonium states using 2010 collected data (35 pb$^{-1}$). The measurement of the production cross sections of charmonium and bottomonium states 
($\psi(2S)$, $\Upsilon(1S)$, $\chi_c$) are useful to test theoretical models. For the exotic states, the measurement of the $X(3872)$ mass and cross section has been
performed and the CDF narrow $X(4140)$ has been studied using a dataset of 376 pb$^{-1}$ and its existence not confirmed. 
LHCb has a very high $J/\psi$ statistic in 2011 data that will allow to have a lot of new results in the near future.

\bigskip 

\end{document}